\documentstyle[12pt,amssymb,amsmath,graphicx]{article}

\advance \topmargin by -\headheight
\advance \topmargin by -\headsep

\evensidemargin \oddsidemargin
\newcommand{\beq}{\begin{equation}}
\newcommand{\eeq}{\end{equation}}

\newcommand{\bea}{\begin{eqnarray}}
\newcommand{\eea}{\end{eqnarray}}

\def\CC{{\mathchoice
{\rm C\mkern-8mu\vrule height1.45ex depth-.05ex
width.05em\mkern9mu\kern-.05em}
{\rm C\mkern-8mu\vrule height1.45ex depth-.05ex
width.05em\mkern9mu\kern-.05em}
{\rm C\mkern-8mu\vrule height1ex depth-.07ex
width.035em\mkern9mu\kern-.035em}
{\rm C\mkern-8mu\vrule height.65ex depth-.1ex
width.025em\mkern8mu\kern-.025em}}}

\def\RR{{\rm I\kern-1.6pt {\rm R}}}

\def\ZZ{{\rm Z}\kern-3.8pt {\rm Z} \kern2pt}
\def\IB{\relax{\rm I\kern-.18em B}}
\def\ID{\relax{\rm I\kern-.18em D}}
\def\II{\relax{\rm I\kern-.18em I}}
\def\IP{\relax{\rm I\kern-.18em P}}

\begin{document}

\begin{titlepage}

\begin{center} \Large \bf IIB Superfluid Flows

\end{center}

\vskip 0.1truein
\begin{center}
Daniel Are\'an$^{a,b}$

\vspace{0.1in}

\begin{center}
$^a$ {International Centre for Theoretical Physics (ICTP)\\
Strada Costiera 11; I 34014 Trieste, Italy \\}
\vskip 5pt
$^b$ {INFN - Sezione di Trieste\\
Via Bonomea 265; I 34136 Trieste, Italy\\}
\vskip 5pt
{\texttt{arean@sissa.it}}

\end{center}

\end{center}
\vskip.5
truein

\begin{center}
\bf ABSTRACT
\end{center}
We construct holographic superfluid flow solutions in a
five-dimensional theory that arises as a consistent truncation of low energy
type IIB string theory. We then study the phase diagram of these systems
in terms of the temperature and superfluid velocity.
Finally, we construct solutions representing the ground
state of these superfluids for velocities below a critical value. These are charged
anisotropic AdS domain walls that demonstrate the emergence of quantum criticality in the IR.

\vskip5.6truecm
\smallskip
\end{titlepage}
\setcounter{footnote}{0}

\section{Introduction}
The observation \cite{Gubser:2008px} that an electrically charged black hole can develop
scalar hair when its temperature is low enough has given rise to much effort in
the study of holographic models of superfluid phase transitions.
The first realizations \cite{HHH} of the so-called holographic superconductors
were based on the minimal set-up of a charged massive scalar minimally coupled to
Einstein-Maxwell theory: in the presence of a charged black hole of sufficiently low
temperature the scalar condenses breaking the electromagnetic gauge symmetry and thus
realizing an spontaneous breaking of a $U(1)$ global symmetry in the dual theory.
Subsequently, several microscopic embeddings of holographic superconductors have been proposed
in the framework of type IIB string theory \cite{Gubser}, M-theory \cite{Gaunt}, and D7-brane models
\cite{Ammon:2009fe}. We will be interested in the model \cite{Gubser} wich is 
a five dimensional truncation of type IIB containing an abelian gauge field and a charged scalar.

In this note which summarizes the results of the papers \cite{ABKP, ABKP2}
we report on some progress on the study of holographic superfluids
in the framework given by the type IIB truncation \cite{Gubser}.
First, as it was done in \cite{Basu,Herzog} for the early phenomenological models, we will put the superfluid
in motion and consider a scenario where there is a non-vanishing superfluid velocity $\xi$, thus
constructing what we call IIB superflows. Holographically this requires a non-trivial profile for a
spatial component of the gauge field, besides the ever-present temporal component dual
to the charge density (or chemical potential) of the system.
We will then proceed to study the zero temperature of these superflows and see if their
ground state is governed by a quantum critical point characterized by the emergence of conformal
invariance in the IR.
Quantum critical points are expected to be of significance in understanding the ground states
of high-$T_c$ superconductors. Holographic constructions of such quantum critical
(hence zero-temperature) superconductors give rise to domain wall solutions, which
capture the holographic RG flow from a symmetric state in the UV to a symmetry-breaking
vacuum in the IR \cite{GubRoch2,GubRoch,Gaunt}.

\subsection{Summary of results}
\begin{itemize}

\item
We find solutions dual to superfluid flows in 3+1 dimensions. These are five-dimensional
black holes with scalar and vector hair and they exist up to a critical value of the superfluid
velocity where the superfluidity is lost, and, contrary to what happens for other models
(see \cite{Basu,Herzog,Tisza,Daniel2}), the phase transition is always second order.
Moreover, at low temperatures and for low enough velocities we find a universal behaviour of
the system with strong indications of the emergence of quantum criticality at zero temperature.

\item
We construct solutions realizing the (zero temperature) ground state of the IIB superflows for superfluid velocities below a critical value $\xi_c$. These geometries are charged anisotropic AdS-to-AdS domain walls which interpolate between
two AdS in the UV and IR. In the UV the non-zero values of the temporal and (one) spatial components
of the gauge field ($A_t\sim\mu\,,\; A_x\sim A_{x,0}$) correspond to the introduction of a chemical potential
and a superfluid velocity respectively. We are deforming the theory by an operator $\sim \mu\,J^0+A_{x,0}\,J^x$ which does not break the $U(1)$ symmetry. The scalar vanishes in the UV, but its subleading asymptotics
realizes a VEV of the dual operator and thus the $U(1)$ is broken spontaneously.
In the IR the solution flows to the same AdS as found in \cite{GubRoch} for zero superfluid velocity, showing the emergence
of relativistic conformal symmetry. The scalar has a non-zero value in the IR AdS breaking explicitly the $U(1)$
symmetry. These domain walls cease to exist for $\xi>\xi_c$, while superfluid flows at very low temperature still exist for those values.
\end{itemize}

The findings of this letter can be summarized in the phase diagram shown in Figure 1. There we speculate about
the nature of the ground state of the superflows for $\xi>\xi_c$ alluding to the similarity of this region with that
believed to be present in some real-life superconductors when there is an imbalance $\delta\mu$ in the chemical potential between the two populations that form the cooper pairs \cite{comb} (see \cite{holofflo} for a holographic model of these unbalanced superconductors). For high enough $\delta\mu$ the system enters an anisotropic phase at $T=0$ known as FFLO phase which is characterized  by a spontaneous breaking of translational invariance \cite{fflo}.
One could think that the ground state of the superflows for $\xi>\xi_c$ is described by a domain wall that interpolates between AdS in the UV and some generalized Lifshizt geometry that breaks isotropy in the IR.

\begin{figure} [ht]
\begin{center}
\includegraphics[height=0.21\textheight]{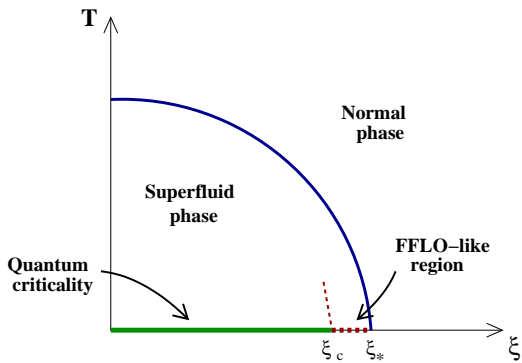}
\label{dxi}
\vskip -8pt
\caption{The (qualitative) phase diagram of holographic superfluids. At zero temperature, a quantum critical point is found for velocities below a critical value, $\xi_c$. Above $\xi_c$ 
the system enters a more anisotropic phase where the deformation induced by $\xi$ affects the RG-flow strongly, and brings the system away from any obvious  
AdS-like IR fixed point.}
\end{center}
\end{figure}

\section{IIB Superflows}
\label{superflows}
The theory we will use is the consistent truncation of low energy type IIB string theory considered in \cite{Gubser}. The action takes the form
\bea
\label{IIBac}
S_{IIB}&=&\int d^5x \sqrt{-g}\bigg[R-{L^2 \over 3}F_{ab}F^{ab}+ {1
\over 4}\left({2L \over 3}\right)^3 \epsilon^{abcde}F_{ab}F_{cd}A_e
+\nonumber
\\
&&- \frac{1}{2}\left((\partial_a \psi)^2 + \sinh^2 \psi(\partial_a
\theta -2 A_a)^2-{6 \over L^2}\cosh^2\left({\psi \over
2}\right)(5-\cosh \psi)\right)\bigg]\,.  \eea
We have set $16 \pi G=1$ and  the conventions are such that $\epsilon^{01234}=1/\sqrt{-g}$, and we have written the charged 
(complex) scalar by splitting the phase and the modulus in the form
$\psi e^{i\theta}$. The Abelian gauge field $A$ is dual to an $R$-symmetry in the boundary field
theory \cite{Gubser} and the scalar field has $R$-charge $R=2$. The general equations of motion that follow from this action (working in the gauge $\theta=0$) can be found in \cite{ABKP}.

We want to construct a fully backreacted hairy black hole solution,
holographically describing a superfluid flow.
To have a charged scalar condense we will need to turn on both the
scalar and the time component of the gauge field in the bulk \cite{Gubser:2008px}.
The superfluid velocity in (say) the $x$-direction is captured by the leading fall-off of
the bulk gauge field component $A_x$ at the boundary, so to describe a superfluid flow
we should allow for a non-trivial profile of $A_x$ in the bulk. To avoid working with partial
differential equations we will restrict to functions depending only on the holographic direction
$r$. Consistency of the Einstein equations
then demands that we choose the following ansatz for the metric, gauge field and scalar:
\bea
ds^2=-{r^2 f(r) \over L^2}dt^2+{L^2 h(r)^2 \over r^2 f(r)} dr^2 -2
C(r) \frac{r^2}{L^2}dt dx+{r^2 \over L^2}B(r) dx^2+{r^2 \over L^2}
dy^2 + {r^2 \over L^2} dz^2\,,\nonumber\\ \nonumber\\
A=A_t(r)\, dt + A_x(r) \,dx\;, \qquad \psi=\psi(r)\,.{\hspace{2.47in}}
\label{suflowansatz}
\eea
Plugging this in the equations of motion we get a set of seven independent equations
for seven unknowns.
These equations exhibit four scaling symmetries which we quote here for convenience:
\bea && t \rightarrow
t/\mbox{a}\;, \qquad f \rightarrow \mbox{a}^2 f\;, \qquad h\rightarrow
\mbox{a} \,h\;, \qquad C \rightarrow \mbox{a} \,C\;, \qquad A_t
\rightarrow \mbox{a} \,A_t\;,
\label{scale1B} \\\nonumber\\
&& x \rightarrow x/\mbox{b}\;, \qquad B \rightarrow \mbox{b}^2 B\;,
\qquad C \rightarrow \mbox{b}\, C\;, \qquad A_x \rightarrow \mbox{b}\,
A_x\;,\label{scale2B}\\\nonumber\\
&& \label{scale3B}(r, t, x, y, z, L) \rightarrow \alpha
(r, t, x, y, z, L)\;, \qquad (A_t, A_x) \rightarrow (A_t, A_x)/\alpha\;,
\\\nonumber\\
\label{scale4B}
&& r \rightarrow \beta r\;, \qquad (t, x, y, z) \rightarrow (t, x, y,
z)/\beta\;, \qquad (A_t, A_x) \rightarrow \beta (A_t, A_x)\;.
\eea
In particular, we will use the symmetries (\ref{scale3B}) and (\ref{scale4B}) to scale the horizon radius $r_H$ and the AdS
scale $L$ to unity.

We will construct our solutions numerically by shooting from the IR. Requiring
regularity at the horizon the solution there can be expanded as
\bea
&&f=f_1^H(r-r_H)+...\,,\quad  h=h_0^H+h_1^H(r-r_H)+...\,,\quad B=B_0^H+B_1^H(r-r_H)+...\label{start}{\qquad}
\nonumber\\\nonumber\\
&&C=C_1^H (r-r_H)+...\,,\quad A_t=A_{t,1}^{H}(r-r_H)+...\,,\quad A_x=A_{x,0}^{H}+A_{x,1}^{H}(r-r_H)+...\nonumber\\
\nonumber\\
&&\psi=\psi_0^H+\psi_1^H(r-r_H)+...\,.
\label{horizonexp}
\eea
The equations of motion impose further constraints on the coefficients of this solution (see  \cite{ABKP}) and in the end
we are left with the following six independent horizon data:
\beq
(h_0^H, B_0^H, C_1^H, A_{t,1}^H, A_{x,0}^H, \psi_0^H)\,.
\label{horizondata5d}
\eeq
We also need to know the asymptotic expansion of our solutions in the UV. An expansion that solves the equations of motion in the UV and is generic enough to match the curves arising from the integration from the horizon is given by
\bea
&&f=h_0^2+\frac{f_4}{r^4}+\frac{f^l_4}{r^4} \log r+...\;, \quad\hspace{1.2cm}
h=h_0+{h_2 \over r^2}+{h_4 \over r^4}+{h^l_4 \over r^4}\log r+...\;,\nonumber\\ \nonumber\\
&&B=B_0+\frac{B_4}{r^4}+\frac{B^l_4}{r^4}\log r+...\;,\quad\hspace{.8cm}
C=C_0+\frac{C_4}{r^4}+\frac{C^l_4}{r^4}\log r+...\;,\nonumber\\ \nonumber\\
&&A_t=A_{t,0}+\frac{A_{t,2}}{r^2}+\frac{A^l_{t,2}}{r^2}\log r+...\;, \quad
A_x=A_{x,0}+{A_{x,2} \over r^2}+\frac{A^l_{x,2}}{r^2}\log r+...\;,\nonumber\\ \nonumber\\
&&\psi=\frac{\psi_1}{r}+\frac{\psi_3}{r^3}+\frac{\psi^l_3}{r^3}\log
r+...\;.\label{UVfalloff}
\eea
As shown in \cite{ABKP} there are further relations between the coefficients and the independent boundary data
can be taken to be $(h_0,f_4, B_0, B_4, C_0, C_4, A_{t,0}, A_{t,2}, A_{x,0}, A_{x,2}, \psi_1,\psi_3)$. To get asymptotically AdS solutions, we must set $B_0, h_0$ to 1 and $C_0, \psi_1$ to zero\footnote{The condition $\psi_1=0$ also ensures, due to the relations between the coefficients \cite{ABKP}, the vanishing of all the logarithmic pieces in (\ref{UVfalloff}).}. The scaling symmetries can be used to accomplish the first two conditions, whereas we need to shoot for the last two. We are therefore left with eight independent
boundary data:
\beq
(f_4, B_4, C_4, A_{t,0}, A_{t,2}, A_{x,0}, A_{x,2}, \psi_3)\,.
\label{UVdata}
\eeq
Notice that there are only six independent pieces of horizon data so we expect to find (if any) a two-parameter family of solutions.

Finally, an important quantity for studying the thermodynamics of the system is of course the superfluid temperature which is given by the black hole Hawking temperature, T.
From the structure of the metric (\ref{suflowansatz}) we easily get 
\beq
\label{temp4d}
T = \frac{r_H^2\, f'(r_H)}{4\pi\,L^2\,h(r_H)}=\frac{1}{4\pi}\left[h_0^H
\left({9 \over 4}+2 \cosh{\psi_0^H}-{\cosh(2\psi_0^H) \over 4}\right)-
\frac{2 (A_{t,1}^H)^2}{9 h_0^H} \right]\,,
\eeq 
where in the second equality we have used the horizon expansion (\ref{horizonexp}), and as explained above we have set $r_H=1$.
\begin{figure}[ht]
\begin{center}
\includegraphics[width=\textwidth]{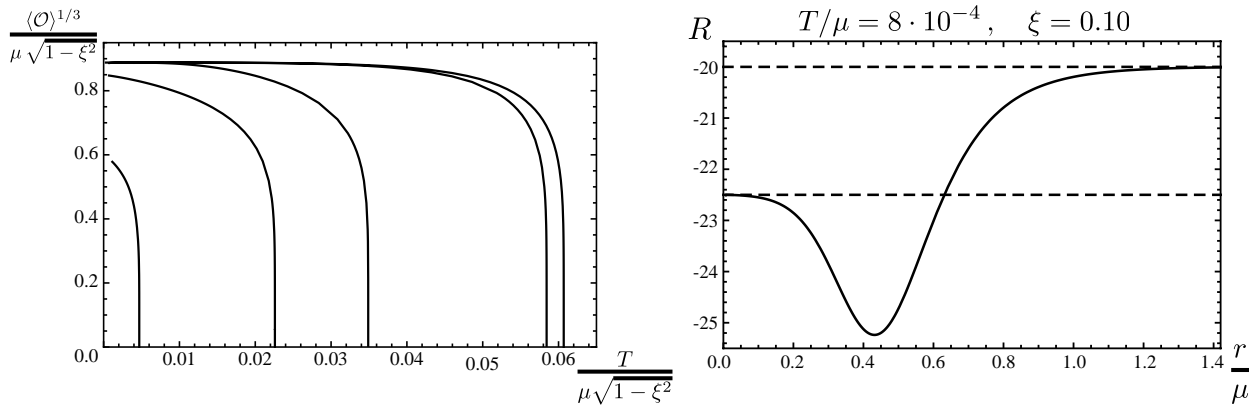}
\label{condplot}
\vskip -8pt
\caption{On the left we show condensate plots for $\xi=0,0.1,0.33,0.4,0.5$ (from right to left).
On the right we plot the Ricci scalar as a function of the radial coordinate near the horizon. The horizontal dashed lines indicate the corresponding values of $R$ for our UV (large $r$) AdS (upper line) and for the IR AdS of \cite{GubRoch}(lower line).}
\end{center}
\end{figure}

The results of our numerical integration are shown in Figure \ref{condplot} where we plot the condensate versus the temperature
for different values of the superfluid velocity. Let us introduce the notation used for the plot:
\beq
\mu \equiv A_{t,0}\,, \qquad {\langle {\cal O}\rangle}\equiv{\sqrt{2}\,\psi_3}\,, \qquad \xi \equiv A_{x,0}/A_{t,0}\,,
\eeq
where $\mu$ is the field theory chemical potential, ${\cal O}$ the (condensing) chiral primary operator, and $\xi$
the superfluid velocity in units of the 
chemical potential. When we work in an ensemble with fixed  chemical
potential the meaningful (dimensionless) quantities relevant for the condensate plot are
$T/\mu$  and $\langle {\cal O}\rangle/\mu^3$.
In constructing the plots we have also rescaled by the (velocity-dependent) factor $\sqrt{1-\xi^2}$, which is
nothing but the relativistic boost factor.
From the form of the curves in Figure \ref{condplot}, it is evident that there is a phase transition to a hairy black hole at low temperatures, and as expected, the critical temperature decreases as the velocity is increased.
To characterize the phase transition one can compare the free energy of the hairy black hole and that
of the normal phase (given by a Reissner-Nordstrom black hole). This has been done in \cite{ABKP} where it
was shown that the phase transition stays second order for all values of the superfluid velocity, up to
our numerical precision.

Finally, from the plot of the condensate one notices that for low superfluid velocities there is a universal behaviour at very low temperatures. Motivated by this fact we check if, at low temperatures, the curvature invariants of our superfluid solutions tend to the values corresponding to the (AdS) quantum critical point found in \cite{GubRoch} for the zero velocity case. Indeed we find so for velocities up to a critical velocity $\xi_c\sim0.37$ (see Figure \ref{condplot}).

\section{Superflows at zero temperature. Anisotropic domain walls}
We will now construct domain wall geometries representing the ground state of IIB superflows for $\xi<\xi_c$.
We will use the same ansatz as in section \ref{superflows}, namely (\ref{suflowansatz}).
In the UV the solution looks as in the previous section. The geometry becomes AdS, and the non-zero values of $A_t$ and $A_x$ correspond to deforming the theory by introducing a chemical potential and a superfluid velocity. The subleading coefficient of the scalar field represents the condensation of the dual operator, which breaks spontaneously the $U(1)$ symmetry and realizes the superfluid phase transition. We expect the UV deformation to trigger an RG flow, and for $\xi<\xi_c$ the solution to flow to the same IR conformal fixed point as in the $\xi=0$ case  \cite{GubRoch}.
Notice that due to the non-trivial profile of $A_x$ these domain walls will be anisotropic.
\begin{figure} [h]
\begin{center}
\includegraphics[width=\textwidth]{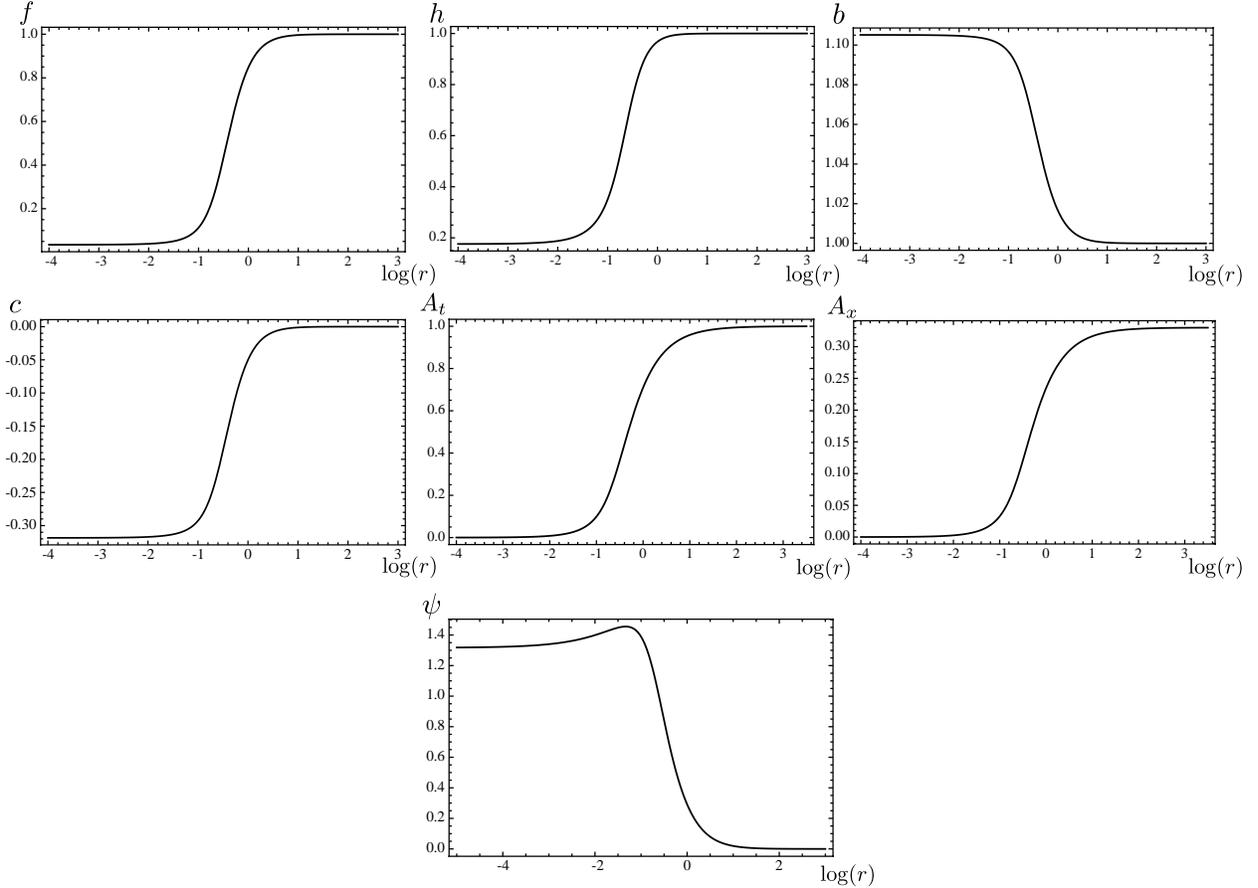}\label{dwalls}
\caption{Plots of the various functions as a function of the radial coordinate for $\xi=0.33$. A logarithmic radial coordinate has been chosen to illustrate that the solution is a domain wall. For other values of the superfluid velocity in the range $0<\xi<\xi_c$ the nature of the curves is similar.}
\end{center}
\end{figure}

We will proceed as before and integrate numerically the equations of motion (see \cite{ABKP}) from the IR up to the boundary. Since we look for solutions that flow in the IR to the symmetry-breaking vacuum found in \cite{GubRoch} we have to find an IR expansion around that vacuum that solves our equations of motion. It reads:
\bea
&&\psi(r)=\operatorname{ArcCosh} 2 + \psi_{1,0} \ x+\psi_{0,2} \ y^2+ ...\;,\quad
x\equiv r^\alpha\;,\quad y\equiv r\;,\quad  \alpha=2 \sqrt{3}-2\,,\qquad\\ \nonumber\\
&&f(r)=\frac{9}{8}h_{0,0}^2+\frac{4\ A_{t(0,2)}^2}{3}\ y^2+...\;,\qquad
h(r)=h_{0,0}+h_{0,2}\ y^2+...\; , \label{h00}\\ \nonumber\\
&&B(r)=B_{0,0}-\frac{4 A_{x(0,2)}^2}{3}\ y^2+...\;, \qquad\; C(r)=C_{0,0}+\frac{4 A_{x(0,2)}A_{t(0,2)}}{3}\ y^2+...\;,
\\ \nonumber\\
&&A_t(r)=A_{t(0,2)}\ y^2+...\;, \qquad\qquad\quad\;\;  A_x(r)=A_{x(0,2)}\  y^2+...\;,
\eea
where not all the coefficients are independent \cite{ABKP2}. As before, there are only six independent quantities
in the IR: $(h_{0,0}\,,\; B_{0,0}\,,\; C_{0,0}\,,\; A_{t(0,2)}\,,\; A_{x(0,2)}\,,\; \psi_{1,0} )$. Our strategy in constructing the solutions will be to pick numerical values for these IR data, and integrate the equations of motion up to some very large $r$ corresponding to the UV. There, a expansion generic enough for our purposes is again (\ref{UVfalloff}). 
To get asymptotically AdS boundary conditions in the UV, we need to set $B_0=1=h_0$, and $C_0=0=\psi_1$. The former two conditions can be accomplished via the rescalings (\ref{scale1B}, \ref{scale2B}), while a shooting technique is required for the latter two. This gives rise to the eight independent boundary quantities (\ref{UVdata}).
We will also use the rescaling (\ref{scale4B}) to set the leading piece of $A_t$ at the boundary (namely $A_{t,0}\equiv \mu$) to unity\footnote{In \cite{GubRoch} this scaling was used to set the horizon datum $A_{t(0,2)}=1$. We prefer instead to set $A_{t,0}=1$ at the boundary: this corresponds to working in a fixed chemical potential ensemble in the gauge theory. In the finite temperature case of section \ref{superflows} this scaling shifts the horizon radius and therefore effectively introduces a new parameter, the temperature.}. 
Once we fix the chemical potential to one, for any given superfluid velocity $\xi$ (before the rescaling of the solution that sets $A_{t,0}$ to unity, $\xi$ is given by $A_{x,0}/A_{t,0}$)  the number of independent parameters at the boundary is six, which is the same as the number of horizon data. Therefore, we expect to find at most discretely many domain wall solutions for any given superfluid velocity. As discussed in \cite{GubRoch} we will look for the solution where the radial profile of the scalar field has the least number of nodes.

The strategy we have described provides the domain wall solution we were seeking.
We present some plots of the various functions for a selected value of the velocity in Figure \ref{dwalls}. 
Plots for other velocities are qualitatively similar. We find that solutions exist for velocities $0\leq\xi<\xi_c$ with $\xi_c \approx 0.374$, which is consistent with the expectations of section \ref{superflows}.
There we have seen that superflows exist at very low temperature up to a higher $\xi_*\sim0.5$. This leaves open the question of what is the ground state of the IIB superflows for $\xi>\xi_*$ about which some speculations were offered in the introduction of this note.

\section*{Acknowledgments}

I would like to thank my collaborators Matteo Bertolini, Chethan Krishnan and Tom\'{a}\v{s} Proch\'{a}zka,
and  the organizers of the XVII European Workshop on String Theory (Padova, September 2011) where this work was presented.
Finally, thanks to the Front of Galician-speaking Scientists for unconditional support.

\end{document}